\documentstyle[12pt]{article}

\begin{document}

\begin {center}
{\large \bf Can gravity make the Higgs particle light?}
\vskip 2cm
{\bf J. J. van der Bij }\\ \bigskip Albert Ludwigs Universit\"at Freiburg\\
Fakult\"at f\"ur Physik \\ H. Herder Strasse 3\\
79104 Freiburg i. Br.\\
\end{center}

\bigskip
\begin{abstract}
The spontaneous symmetry breaking theory of gravity is examined,
assuming that the vacuum expectation value of the standard model Higgs
is also responsible for the generation of the Planck mass.
In this model the physical Higgs couples only with gravitational
strength to matter.
At presently accessible energies the theory is indistinguishable
from the standard model without Higgs boson and is in agreement
with all existing data.At energies above the Fermi scale new dynamics should
occur.
\end{abstract}
\bigskip
\bigskip
\bigskip
\bigskip

The standard model is in agreement with existing data.
However it leaves many questions unanswered, in particular
with regard to the masses of particles. The mass spectrum of
particles is completely ununderstood. In the standard model all
masses arise from the coupling to the Higgs sector. It is therefore
the Higgs sector, where the dynamics of mass generation is hidden. Another
problem
of the Higgs sector is of a more theoretical but possibly more
fundamental nature. It is the so-called naturalness problem.
The corrections to the Higgs mass are quadratically divergent.
Therefore one expects the Higgs to be as heavy as the cut-off
of the theory, although within the standard model the Higgs mass is supposed
to be close to the Fermi scale.
Suggested solutions to the naturalness
problem are technicolor \cite{far}, supersymmetry \cite{nil} and heavy quark
condensates \cite{lin}.

In technicolor there is no fundamental Higgs and the only interactions are
gauge interactions. Therefore all radiative corrections are only
logarithmically divergent. Unfortunately no working model has been found
sofar.

 In supersymmetry there are no quadratic divergences in the
theory due to cancellations between boson and fermion loops.
A minimal supersymmetric standard model exists that is not in conflict
with present data . Unfortunately the model essentially doubles
the present spectrum of particles replacing Bose (Fermi) degrees
of freedom with Fermi (Bose) degrees of freedom. About the nature
of the Higgs sector itself little is said.

A third attempt is to cancel the quadratic divergences within the
standard model itself. This gives rise to relations between top
and Higgs mass. However these cancellations can
only be valid at one scale, as the top quark feels
QCD corrections and the Higgs not. This leads  to the idea that
the Higgs is a $\bar tt$ condensate. Also here no satisfactory model
seems to exist.

In this letter I will consider a fourth logical possibility.
Namely maybe the cut-off of the theory is not so large and
of the order of the weak scale, but strong interactions appear.
All explanations given above leave gravity out of the picture.
While historically it has been emphasized that gravity may not
play a role in elementary particle physics because of the
large discrepancy between the Fermi scale and the Planck scale,
this may be a mistake.
To argue that gravity may play a role there are a few points I
want to emphasize.
First there is the anomaly. There is an anomaly in the gauge currents
due to the coupling to gravity, that has to cancel to preserve
gauge invariance. This leads to the condition Tr(Y) = 0, which is
indeed satisfied. Therefore it appears that gravity is sensitive to the
gauge structure of the standard model.

 A second point is that
the Higgs and the graviton have some similar characteristics. The Higgs
is coupled universally to the mass of particles and the graviton
to energy-momentum. So at least at low energy one might conjecture
a relation between the Higgs sector and gravity. The fact that the
Higgs potential generically generates a cosmological constant
may also be an indication for a connection.

 As mentioned above
the problem establishing such a relation is the large discrepancy
between the Fermi scale and the Planck scale. I will try to adress
this problem in a model with spontaneous symmetry breakdown.
Within models of spontaneous symmetry breaking there are essentially
two ways to generate a large difference in mass scales.
The standard way is to assume that there are several fields having
very different vacuum expectation values, but similar values of
coupling constants. This way the theory stays perturbative
at all scales. The other possibility is to have fields with
similar vacuum expectation values, but widely different coupling
constants. Models of this type with singlet Higgs fields are
considered in \cite{hill}, giving strong effects in W-boson interactions.
Including gravity a model of this type is given by the spontaneous
symmetry breaking theory of gravity . In this model the Planck constant
is generated through the vacuum expectation value of a Higgs
field $\phi$ that is coupled to gravity via a $\xi \phi^2 R$ term. To my
knowledge the first paper of this type is \cite{fuji}. The subject became
popular through the article \cite{zee}. A review is given in \cite{adler}.
In the literature the fields $\phi$ that have been discussed are
typically singlet fields or grand unified fields, with a large
vacuum expectation value of the order of the Planck mass, so that the
coupling constant $\xi$ is relatively small. Often a bare $R$ term in
the action is  assumed, so that the non-minimal
term is a small correction. At low energies these models are indistinguishable
from the standard model.

In this letter I will consider the case
that the $\phi$ field is the Higgs field of the standard model itself.
This possibility appears not to have been considered before because of the
large value of $\xi$ that is needed.
The Lagrangian of the model is given by:
$$ {\cal L} = \sqrt{g}\bigl ( \xi \Phi^{+} \Phi R -\frac {1}{2} g^{\mu \nu}
  (D_{\mu} \Phi)^{+}(D_{\nu}\Phi) -V(\Phi^{+} \Phi) - \frac {1}{4}
F_{\mu \nu} F^{\mu \nu} \bigr )\eqno(1)$$
If we take $\Phi$ to be the standard model Higgs field the coupling
constant $\xi$ is very large. In order to give the Einstein Lagrangian
after spontaneous symmetry breakdown one needs $\xi = \frac {\kappa^2}
{v^2} = \frac {1}{16 \pi G_N v^2} \approx O(\frac {m^2_{Pl}}{v^2})$.
Due to this very large value of $\xi$ the physical content of the theory
is not quite manifest. The physical content can be made manifest
by the Weyl rescaling $
g_{\mu \nu} \rightarrow \frac{\kappa^2}{\xi v^2}
g_{\mu \nu}$.
Keeping only the Higgs and the gravity part the Lagrangian  becomes:
$$ {\cal L} = \sqrt{g}\bigl ( \kappa^2 R -\frac{3}{2}\frac {\xi v^2}
{\vert \Phi \vert^4}
(\partial_{\mu}
\vert \Phi \vert ^2)(\partial^{\mu} \vert \Phi \vert ^2)
-\frac {1}{2} \frac {v^2}{\vert \Phi \vert^2} (D_{\mu} \Phi^{+})
(D^{\mu} \Phi) - \frac {v^4}{\vert \Phi \vert^4}  V(\vert \Phi \vert^2)
\bigr )
\eqno(2) $$

The gravity part of the Lagrangian has now become of the canonical form.
The Higgs part is a modified form of the standard model. As potential
we take $ V(\Phi) = \frac {1}{8} \lambda (\Phi^+ \Phi - v^2)^2$
To analyze the physical content of the theory one takes the unitary
gauge $\Phi = {0 \choose v + \sigma }$. Expanding around the minimum
one gets for the quadratic part of the Higgs Lagrangian
$${\cal L}_{Higgs} = -\frac{1}{2} (1 + 12 \xi) (\partial_{\mu} \sigma)
(\partial^{\mu} \sigma) - \frac {1}{2} \lambda v^2 \sigma^2 \eqno(3)$$
Therefore there is a wave function renormalization of the Higgs
field by a factor $1/ \sqrt {1 + 12 \xi}$. As a result the
effective coupling of the Higgs field to matter becomes of gravitational
strength $O( m / m_{Pl})$.
The mass of the Higgs particle itself is given by
$m_H^2 = \frac {\lambda v^2}{1 + 12 \xi}$. Because the coupling of
the Higgs to matter is of gravitational strength the Higgs becomes
essentially a stable particle. This could have some cosmological
consequences and remnant Higgs particles could still be present.
For $\lambda \approx O(1)$
$m_H$ becomes very small and this will result in a contribution to the
gravitational force with a range of $1.9 cm/ \sqrt{\lambda}$.
The coupling strength to matter is given by
$$(\frac {1}{1+12\xi})^{1/2} ( g_q \bar qq + \frac {\alpha_s}{2\pi}
G_{\mu \nu} G^{\mu \nu}) \eqno(4)$$
The Higgs particle behaves similar to the cosmon  \cite{pec}.

When one assumes that
$\lambda \approx O(\xi)$, $m_H$ becomes of
the order of the electroweak scale. Because the Higgs coupling
has been reduced to gravitational strength, the Higgs particle
effectively decouples from the theory and one is left with
the massive Yang-Mills theory. The massive Yang-Mills theory
is in agreement with present experiments even though it is
non renormalizable, because the cut-off dependence is only logarithmic
(Veltmans theorem) \cite{velt}. Therefore if the cut-off of the theory is
around the Fermi scale, the present experiments at LEP are not sensitive
to such effects.

The way the Higgs is removed from the theory here
by making its coupling to matter small is to be contrasted with the usual way
where the mass of the Higgs is taken to be large. As is well known
in such a limit the theory becomes equivalent to a gauged non-linear
$\sigma$ model. The longitudinal W-bosons correspond to the pions
of the theory. The scattering amplitude of the pions should go
to zero at low energy. This is indeed the case for the Lagrangian [2].
As an example one has for the 4-point scattering amplitude
the following formula in lowest order.
$${\cal M} = \frac{1}{v^2} \bigl [ \frac{(12 \xi k^2 + \lambda v^2)^2}
{(1+12 \xi)k^2 + \lambda v^2}- (12 \xi k^2 + \lambda v^2) \bigr]
\delta_{ab} \delta_{cd} + {\rm permutations} \eqno(5)$$
which behaves as $- \frac {k^2}{v^2} \delta_{ab} \delta_{cd}$
for $k \rightarrow 0$, so that indeed an Adler zero is present.

Therefore it has been shown that at energies below the Fermi scale the model is
equivalent with the standard model without Higgs particle.
This leaves open the question what happens at energies above the Fermi scale.
At these energies the non-renormalizability of the theory plays a role
and strong interactions should be present. What the nature of these
strong interactions should be is not clear. As a first approximation
one would expect some form of chiral perturbation theory to be
valid. One could also conjecture that gravity might play a role.
This would be more natural however in grand unified theories, where
$\xi$ is not so large. The model makes clear that the method of mass generation
and the question of renormalizability
are two separate issues. The necessary strong interactions should give
interesting physics at the planned future colliders.
\\

{\bf Acknowledgment}\\
I would like to thank NIKHEF-H and the Nuclear Physics Institute of
Moscow State University, where part of this work was done, for their
hospitality.\\

\end{document}